\documentclass[twocolumn,showpacs,showkeys,preprintnumbers,amsmath,amssymb,aps,nofootXinbib,floatfix]{revtex4}
\usepackage{color}
\usepackage{indentfirst}
\usepackage{graphicx}
\usepackage{multirow}
      \def\di{\displaystyle}
      
      \def\bS{{\bf S}}
      \def\bl{{\bf l}}
      \def\bp{{\bf p}}
      \def\bq{{\bf q}}
      \def\br{{\bf r}}

      \def\F{{\cal F}}
      \def\H{{\cal H}}

      \def\L{{\cal L}}
      
      \def\P{{\cal P}}
      \def\R{{\cal R}}

\begin{document}

\title{
Spin and orbital scissors modes in $^{166}$Er}

\author{ E.B. Balbutsev\email{balbuts@theor.jinr.ru},
I.V. Molodtsova\email{molod@theor.jinr.ru}}
\affiliation{Bogoliubov Laboratory of Theoretical Physics, Joint Institute for Nuclear Research, 141980 Dubna, Russia}

\begin{abstract}
Recently, low-energy dipole excitations in $^{166}$Er were investigated via nuclear resonance fluorescence [T. Shizuma {\it et al}, Phys. Rev. C {\bf 113}, 044325 (2026)]. The magnetic dipole strength associated with the nuclear scissors mode was extracted for excitation energies between 2.2 and 3.5 MeV. It was found that the $M1$ strength distribution is separated into two groups. We interpret this splitting as caused by the spin degrees of freedom instead of the
nucleus nonaxiality suggested by authors.

\end{abstract}


\maketitle

{\it Introduction.}

\begin{figure}[h!]
\centering
\includegraphics[width=\columnwidth]{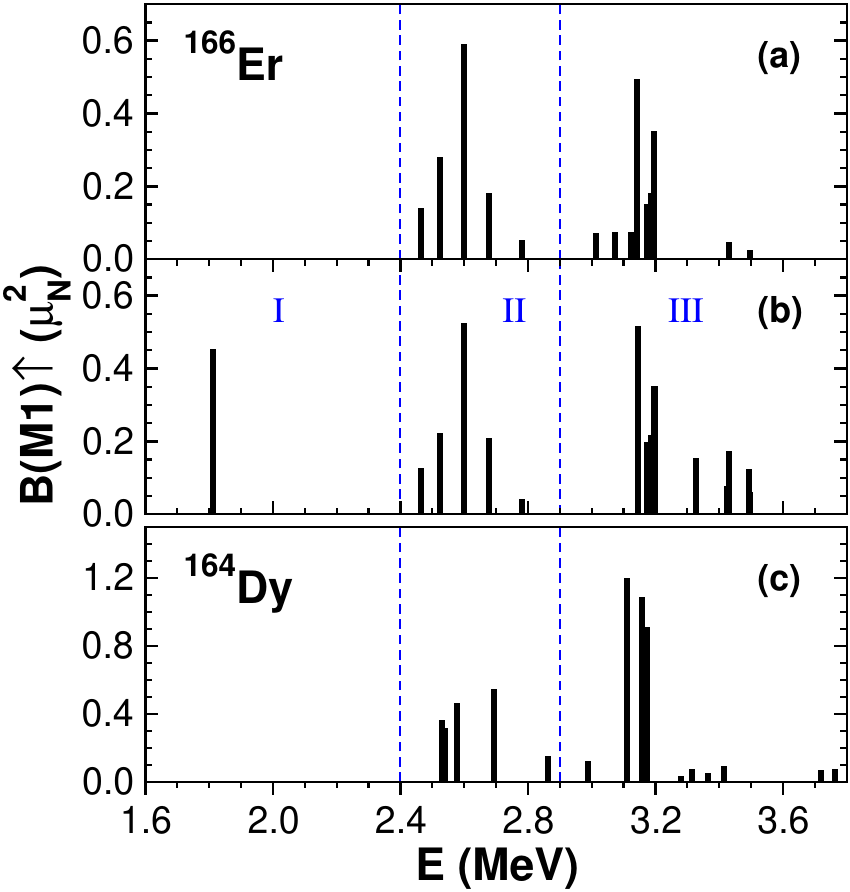}
\caption{ $M1$ strength distribution observed in $^{166}$Er and $^{164}$Dy via the NRF experiments: (a) -- the results of Ref.~\cite{Shizuma26}, (b) -- the results of Ref.~\cite{Maser96},
(c) -- the results of Ref.~\cite{Margraf}.
The dashed lines indicate the boundaries of the energy intervals for $M1$ strength  summation adopted in the paper~\cite{Shizuma26}.}
\label{fig1}\end{figure}

\begin{figure}[h!]
\includegraphics[width=\columnwidth]{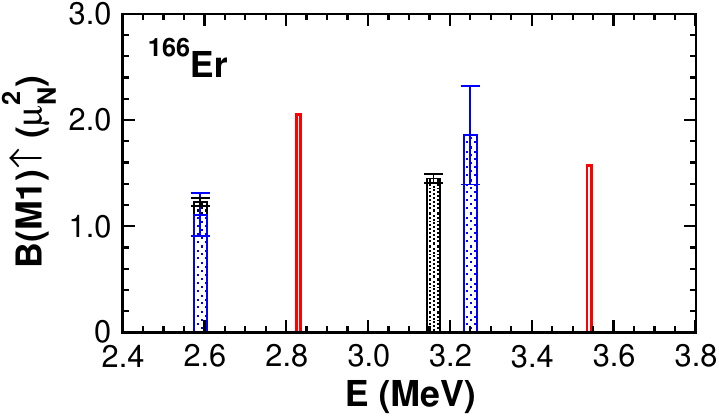}
\caption{The observed and calculated $M1$ strength separately  for intervals II and III, defined in Fig.~\ref{fig1}: black rectangles are the strengths calculated in~\cite{Shizuma26};  blue rectangles are the results from~\cite{Maser96}.
Red rectangles correspond to WFM calculations, see Table~\ref{tab2}.}
\label{fig2}
\end{figure}

The present Rapid Communication is motivated by recent results of studying the scissors mode in $^{166}$Er via nuclear resonance fluorescence, reported in Ref.~\cite{Shizuma26}.
The authors reproduced almost exactly the results of the paper~\cite{Maser96} in the energy interval between 2.4 and 3.5 MeV (see Fig.~\ref{fig1}). The found spectrum of $1^+$ excitations is manifectly splitted in two groups with energy centroids 2.59 MeV and 3.16 MeV with the
strengths $\sum B(M1)=1.23\ \mu_N^2$ and $\sum B(M1)=1.45\ \mu_N^2$, respectively (see Fig.~\ref{fig2}).
The authors suggest to explain this splitting by the nonaxiality of the nucleus. However such explanation looks not very convincing, because $^{166}$Er does not belong to the group of $\gamma$-soft nuclei.

 We want to remind about the similar situation with the interpretation of the results of Oslo group. Guttormsen {\it et al}~\cite{Oslo} have studied deuteron and $^3$He-induced reactions on $^{232}$Th and found in the residual nuclei $^{231,232,233}$Th and
$^{232,233}$Pa ``an unexpectedly strong integrated strength of $B(M1)=11-15\ \mu_N^2$ in the $E_\gamma=1.0-3.5$ MeV region''. The $B(M1)$ force in most nuclei shows evident splitting into two lorentzians. ``Typically, the experimental splitting is $\Delta\omega\sim0.7$ MeV, and the ratio of the strengths between the lower ($B_L$) and upper ($B_U$) resonance components is $B_L/B_U\sim2$.''
The authors have tried to explain the splitting by a $\gamma$-deformation. To describe the observed value of $\Delta\omega$ the deformation $\gamma\sim 15$\textdegree\   is required, which leads to the ratio $B_L/B_U\sim 0.7$ in an obvious contradiction with experiment. The authors conclude that ``the splitting may be due to other mechanism''.

Such other mechanism, the nuclear spin scissors mode,
was suggested and developed in the series of  papers~\cite{BaMoSc,BaMoPRC13,BaMoPRC15,BaMoPRC18,BaMoPRC22,BaMoIJMP24,BaMoEPJ24} where
the Wigner Function Moments (WFM) method was used to solve the
time dependent Hartree-Fock-Bogoliubov (TDHFB) equations.

{\it Formalism.}

TDHFB equations in matrix formulation \cite{Solov,Ring} are
\begin{equation}
i\hbar\dot\R=[\H,\R]
\label{tHFB}
\end{equation}
with
\begin{equation}
\R={\hat\rho\qquad-\hat\kappa\choose-\hat\kappa^{\dagger}\;\;1-\hat\rho^*},
\quad\H={\hat
h\quad\;\;\hat\Delta\choose\hat\Delta^{\dagger}\quad-\hat h^*}
\end{equation}
The normal density matrix $\hat \rho$ and Hamiltonian $\hat h$ are
hermitian whereas the anomalous density $\hat\kappa$ and the pairing
gap $\hat\Delta$ are skew symmetric: $\hat\kappa^{\dagger}=-\hat\kappa^*$,
$\hat\Delta^{\dagger}=-\hat\Delta^*$.
The detailed form of the TDHFB equations is
\begin{eqnarray}
&& i\hbar\dot{\hat\rho} =\hat h\hat\rho -\hat\rho\hat h
-\hat\Delta \hat\kappa ^{\dagger}+\hat\kappa \hat\Delta^\dagger,
\nonumber\\
&&-i\hbar\dot{\hat\rho}^*=\hat h^*\hat\rho ^*-\hat\rho ^*\hat h^*
-\hat\Delta^\dagger\hat\kappa +\hat\kappa^\dagger\hat\Delta ,
\nonumber\\
&&-i\hbar\dot{\hat\kappa} =-\hat h\hat\kappa -\hat\kappa \hat h^*+\hat\Delta
-\hat\Delta \hat\rho ^*-\hat\rho \hat\Delta ,
\nonumber\\
&&-i\hbar\dot{\hat\kappa}^\dagger=\hat h^*\hat\kappa^\dagger
+\hat\kappa^\dagger\hat h-\hat\Delta^\dagger
+\hat\Delta^\dagger\hat\rho +\hat\rho^*\hat\Delta^\dagger .
\label{HFB}
\end{eqnarray}

Let us consider their matrix form in coordinate
space keeping all spin indices $s, s'$:
$\langle \br,s|\hat\rho|\br',s'\rangle$,
$\langle \br,s|\hat\kappa|\br',s'\rangle$, etc.
 We do not specify the isospin indices in order to make
formulae more transparent.
After introducing the more compact notation
$\langle \br,s|\hat X|\br',s'\rangle =X_{rr'}^{ss'}$
the set of equations (\ref{HFB}) with specified spin indices reads
\begin{widetext}
\begin{eqnarray}
&&i\hbar\dot{\rho}_{rr''}^{\uparrow\uparrow} =
\int\!d^3r'(
 h_{rr'}^{\uparrow\uparrow}\rho_{r'r''}^{\uparrow\uparrow}
-\rho_{rr'}^{\uparrow\uparrow} h_{r'r''}^{\uparrow\uparrow}
+\hat h_{rr'}^{\uparrow\downarrow}\rho_{r'r''}^{\downarrow\uparrow}
-\rho_{rr'}^{\uparrow\downarrow} h_{r'r''}^{\downarrow\uparrow}
-\Delta_{rr'}^{\uparrow\downarrow}{\kappa^{\dagger}}_{r'r''}^{\downarrow\uparrow}
+\kappa_{rr'}^{\uparrow\downarrow}{\Delta^{\dagger}}_{r'r''}^{\downarrow\uparrow}),
\nonumber\\
&&i\hbar\dot{\rho}_{rr''}^{\uparrow\downarrow} =
\int\!d^3r'(
 h_{rr'}^{\uparrow\uparrow}\rho_{r'r''}^{\uparrow\downarrow}
-\rho_{rr'}^{\uparrow\uparrow} h_{r'r''}^{\uparrow\downarrow}
+\hat h_{rr'}^{\uparrow\downarrow}\rho_{r'r''}^{\downarrow\downarrow}
-\rho_{rr'}^{\uparrow\downarrow} h_{r'r''}^{\downarrow\downarrow}),
\nonumber\\
&&i\hbar\dot{\rho}_{rr''}^{\downarrow\uparrow} =
\int\!d^3r'(
 h_{rr'}^{\downarrow\uparrow}\rho_{r'r''}^{\uparrow\uparrow}
-\rho_{rr'}^{\downarrow\uparrow} h_{r'r''}^{\uparrow\uparrow}
+\hat h_{rr'}^{\downarrow\downarrow}\rho_{r'r''}^{\downarrow\uparrow}
-\rho_{rr'}^{\downarrow\downarrow} h_{r'r''}^{\downarrow\uparrow}),
\nonumber\\
&&i\hbar\dot{\rho}_{rr''}^{\downarrow\downarrow} =
\int\!d^3r'(
 h_{rr'}^{\downarrow\uparrow}\rho_{r'r''}^{\uparrow\downarrow}
-\rho_{rr'}^{\downarrow\uparrow} h_{r'r''}^{\uparrow\downarrow}
+\hat h_{rr'}^{\downarrow\downarrow}\rho_{r'r''}^{\downarrow\downarrow}
-\rho_{rr'}^{\downarrow\downarrow} h_{r'r''}^{\downarrow\downarrow}
-\Delta_{rr'}^{\downarrow\uparrow}{\kappa^{\dagger}}_{r'r''}^{\uparrow\downarrow}
+\kappa_{rr'}^{\downarrow\uparrow}{\Delta^{\dagger}}_{r'r''}^{\uparrow\downarrow}),
\nonumber\\
&&i\hbar\dot{\kappa}_{rr''}^{\uparrow\downarrow} = -\hat\Delta_{rr''}^{\uparrow\downarrow}
+\int\!d^3r'\left(
 h_{rr'}^{\uparrow\uparrow}\kappa_{r'r''}^{\uparrow\downarrow}
+\kappa_{rr'}^{\uparrow\downarrow} {h^*}_{r'r''}^{\downarrow\downarrow}
+\Delta_{rr'}^{\uparrow\downarrow}{\rho^*}_{r'r''}^{\downarrow\downarrow}
+\rho_{rr'}^{\uparrow\uparrow}\Delta_{r'r''}^{\uparrow\downarrow}
\right),
\nonumber\\
&&i\hbar\dot{\kappa}_{rr''}^{\downarrow\uparrow} = -\hat\Delta_{rr''}^{\downarrow\uparrow}
+\int\!d^3r'\left(
 h_{rr'}^{\downarrow\downarrow}\kappa_{r'r''}^{\downarrow\uparrow}
+\kappa_{rr'}^{\downarrow\uparrow} {h^*}_{r'r''}^{\uparrow\uparrow}
+\Delta_{rr'}^{\downarrow\uparrow}{\rho^*}_{r'r''}^{\uparrow\uparrow}
+\rho_{rr'}^{\downarrow\downarrow}\Delta_{r'r''}^{\downarrow\uparrow}
\right).
\label{HFsp}
\end{eqnarray}
This set of equations must be complemented by the complex conjugated equations.
Writing these equations we neglected the diagonal in spin matrix elements
of the anomalous density:
$\kappa_{rr'}^{ss}$ and $\Delta_{rr'}^{ss}$. It was shown in~\cite{BaMoPRC15}
that such an approximation works very well in the case of monopole pairing
considered here.

 The microscopic Hamiltonian of the model, harmonic oscillator with
spin orbit potential plus separable quadrupole-quadrupole and
spin-spin residual interactions is given by
\begin{eqnarray}
\label{Ham}
 \hat h=\sum\limits_{i=1}^A\left[\frac{\hat\bp_i^2}{2m}+\frac{1}{2}m\omega^2\br_i^2
-\eta\hat \bl_i\hat \bS_i\right]+H_{qq}+H_{ss},
\end{eqnarray}
with
\begin{eqnarray}
\label{Hqq}
&& H_{qq}=\!
\sum_{\mu=-2}^{2}(-1)^{\mu}
\left\{\bar{\kappa}
 \sum\limits_i^Z\!\sum\limits_j^N
+\frac{\kappa}{2}
\left[\sum\limits_{i,j(i\neq j)}^{Z}
+\sum\limits_{i,j(i\neq j)}^{N}
\right]
\right\}
q_{2-\mu}(\br_i)q_{2\mu}(\br_j)
,
\\
\label{Hss}
&&H_{ss}=\!
\sum_{\mu=-1}^{1}(-1)^{\mu}
\left\{\bar{\chi}
 \sum\limits_i^Z\!\sum\limits_j^N
+\frac{\chi}{2}
\left[
\sum\limits_{i,j(i\neq j)}^{Z}
+\sum\limits_{i,j(i\neq j)}^{N}
\right]
\right\}
\hat S_{-\mu}(i)\hat S_{\mu}(j)
\,\delta(\br_i-\br_j),
\end{eqnarray}
where
$\displaystyle q_{2\mu}=\sqrt{16\pi/5}\,r^2Y_{2\mu}=
\sqrt{6}\{r\otimes r\}_{\lambda\mu},\,
\{r\otimes r\}_{\lambda\mu}=\sum_{\sigma,\nu}
C_{1\sigma,1\nu}^{\lambda\mu}r_{\sigma}r_{\nu},\,
C_{1\sigma,1\nu}^{\lambda\mu}$
is the Clebsch-Gordan coefficient,
cyclic coordinates $r_{-1}, r_0, r_1$ are defined in~\cite{Var},
$N$ and $Z$ are the numbers of neutrons and protons.
$\hat S_{\mu}$ are spin matrices~\cite{Var}.
\end{widetext}

Equations (\ref{HFsp}) will be solved by the WFM method. To this  end we rewrite them with
the help of the Wigner transformation \cite{Ring}.
So, instead of four
$(s,s'=\pm\frac{1}{2})$ matrix elements of
the density matrix $\rho_{rr'}^{ss'}$ and two matrix elements
$\kappa_{rr'}^{\uparrow\downarrow}$, $\kappa_{rr'}^{\downarrow\uparrow}$
we will work with four Wigner functions $f^{ss'}(\br,\bp)$ and two phase space
distributions $\kappa^{ss'}(\br,\bp)$, that is more convenient for WFM method
(see \cite{BaMoPRC13, BaMoPRC15} for details).

Following the papers \cite{BaMoSc, BaMoPRC13, BaMoPRC15} in the next step we write above equations in terms of spin-scalar
\mbox{$f^+=f^{\uparrow\uparrow}+ f^{\downarrow\downarrow}$}
and spin-vector
\mbox{$f^-=f^{\uparrow\uparrow}- f^{\downarrow\downarrow}$}
functions.
As a result, we obtain a set of twelve equations, which is solved by the method of moments in a small amplitude
approximation. To this end
all functions $f(\br,\bp,t)$ and $\kappa(\br,\bp,t)$ are divided into equilibrium part
and deviation (variation): $f(\br,\bp,t)=f(\br,\bp)_{eq}+\delta f(\br,\bp,t)$,
\mbox{$\kappa(\br,\bp,t)=\kappa(\br,\bp)_{eq}+\delta \kappa(\br,\bp,t)$}.
Then equations are linearized neglecting quadratic
in $\delta f$ and $\delta \kappa$ terms~\cite{BaMoPRC15}.
 Integrating the transformed equations over phase space
with the weights
\begin{equation}
\{r\otimes p\}_{\lambda\mu},\,\{r\otimes r\}_{\lambda\mu},\,
\{p\otimes p\}_{\lambda\mu}, \mbox{ and } 1
\nonumber
\end{equation}
one gets the set of coupled dynamic equations for the
following collective variables (see Ref.~\cite{BaMoPRC22}):
\begin{eqnarray}
&&\L^{\tau\varsigma}_{\lambda\mu}(t)=\int\! d(\bp,\br) \{r\otimes p\}_{\lambda\mu}
\delta f^{\tau\varsigma}(\br,\bp,t),
\nonumber\\&&
\R^{\tau\varsigma}_{\lambda\mu}(t)=\int\! d(\bp,\br) \{r\otimes r\}_{\lambda\mu}
\delta f^{\tau\varsigma}(\br,\bp,t),
\nonumber\\
&&\P^{\tau\varsigma}_{\lambda\mu}(t)=\int\! d(\bp,\br) \{p\otimes p\}_{\lambda\mu}
\delta f^{\tau\varsigma}(\br,\bp,t),
\nonumber\\&&
\F^{\tau\varsigma}(t)=\int\! d(\bp,\br)
\delta f^{\tau\varsigma}(\br,\bp,t),
\nonumber\\
&&\tilde{\L}^{\tau}_{\lambda\mu}(t)=\int\! d(\bp,\br) \{r\otimes p\}_{\lambda\mu}
\delta \kappa^{\tau}(\br,\bp,t),
\nonumber\\&&
\tilde{\R}^{\tau}_{\lambda\mu}(t)=\int\! d(\bp,\br) \{r\otimes r\}_{\lambda\mu}
\delta \kappa^{\tau}(\br,\bp,t),
\nonumber\\
&&\tilde{\P}^{\tau}_{\lambda\mu}(t)=\int\! d(\bp,\br) \{p\otimes p\}_{\lambda\mu}
\delta \kappa^{\tau}(\br,\bp,t),
\label{Varis}
\end{eqnarray}
 where
$\tau$  is the isospin index and
$\varsigma\!=+,\,-,\,\uparrow\downarrow,\,\downarrow\uparrow$,
 $\int\! d(\bp,\br)\equiv
(2\pi\hbar)^{-3}\int\! d\br\int\! d\bp$.

{\it Results.}

The solution of equations for variables~(\ref{Varis})
produces fourteen $1^+$ states~\cite{BaMoPRC22}.
Three of them (see Table~\ref{tab1}) correspond to
three types of nuclear scissors mode: the conventional one (rotational oscillations of
neutrons against protons, orbital mode) and two spin modes (the simple spin scissors -
rotational oscillations of ``spin-up'' nucleons with respect of ``spin-down'' ones and the
complicate spin scissors - rotational oscillations of
``spin-up'' protons together with ``spin-down'' neutrons against
``spin-down'' protons together with ``spin-up'' neutrons).

Only two of these excitations fall into the discussed energy interval 2.4 - 3.5 MeV
(see Fig.~\ref{fig2} and Table~\ref{tab1}). As it is seen, the agreement of the theoretical results with
the experimental data is rather reasonable, both the energies and the excitation
probabilities, see also Table~\ref{tab2}.

\begin{table}[h!]
\caption{The calculated energies $E$ (MeV) and excitation probabilities
$B(M1)$ ($\mu_N^2$) of three scissors states in $^{166}$Er.
Here $q$ is the spin quenching factor: $g_s^\tau=q g_s^{\tau\rm free}$; $q=0$ means that the spin part of the external field is omitted, and the $M1$ strength is purely orbital.}
\begin{ruledtabular}\begin{tabular}{cccc}
 & $E_i$ (MeV)  & \multicolumn{2}{c}{$B_i(M1)$ ($\mu_N^2$)}  \\
 \cline{3-4}
 $i$ & &  $q = 0.7$ &  $q = 0$ \\
\hline
1& 2.23 & 1.89 & 0.58 \\
2& 2.83 & 2.05 & 1.34 \\
3& 3.54 & 1.57 & 6.82 \\
\end{tabular}\end{ruledtabular}\label{tab1}
\end{table}

The analysis
of currents (see Ref.~\cite{BaMoPRC22}) in $^{166}$Er shows that the excitation with $E=2.83$ MeV
represents predominantly (57\%) the simple spin scissors with rather big
admixture (29\%) of the conventional scissors and 14\% of the complicate spin scissors.
Therefore this excitation has mainly (71\%) spin nature with 29\% of the orbital
nature and approximately can be counted as the spin excitation. On the other side the excitation with $E=3.54$ MeV
represents predominantly (62\%) the conventional scissors with a rather strong
admixture (31\%) of the complicate spin scissors and a small admixture
(7\%) of the simple spin scissors. Thus, this excitation has mainly (62\%) the orbital nature with 38\% of the spin nature.
A similar result was obtained for $^{164}$Dy~\cite{BaMoPRC22}.

The calculations with and without the spin part of the dipole magnetic operator
show that all three scissors states are equally sensitive to the spin dependent part of the external field
(compare columns 3 and 4 in the Table~\ref{tab1}).
The inclusion of the spin part 
plays in favor of spin scissors.
As one sees, the spin influence is very strong.
Nevertheless, from the above analysis it follows that the 3.54 MeV state should be considered as predominantly orbital.
Similar results are obtained also in the RPA calculations, see for example Ref.~\cite{NesterPRC103}.

It turns out that the experimental situation in $^{164}$Dy follows rather closely to the above described picture~\cite{Margraf} (see Fig.~\ref{fig1}).
Frekers {\it et al} \cite{Frek} have found the spin part of the $M1$-transition probability
$B_{\sigma}(M1)=0.38$ $\mu_N^2$ for the 2.53 MeV state,
$B_{\sigma}(M1)=0.34$ $\mu_N^2$ for the 2.66 MeV state,
i.e. the summed value $\Sigma B_{\sigma}(M1)=0.72$ $\mu_N^2$ for their centroid $E_c=2.59$ MeV
and $B_{\sigma}(M1)=0.50$ $\mu_N^2$ for the 3.14 MeV state.
%
%
The authors of Ref.~\cite{Maser96} did not analyze the resonances on the subject of their spin or orbital nature. Nevertheless it looks quite natural to expect, that the situation in $^{166}$Er will be similar to the one in $^{164}$Dy: the lower resonance has the spin nature and the upper resonance has the orbital nature,
that contradicts to the results of calculations within the framework of the
Quasiparticle Vacua Shell Model (QVSM), reported in the paper~\cite{Shizuma26}.

\begin{widetext}

\begin{table*}[h!]
\caption{The results of NRF experiments and WFM calculations for $^{166}$Er:
energy centroids of lower $E_L$ and upper $E_U$ scissors modes (ScM) with corresponding strengths $B_L$ and $B_U$.
Summation is over regions II (Lower) and III (Upper).
}
\begin{ruledtabular}\begin{tabular}{cccccccccccc}
 \multicolumn{6}{c}{Lower ScM} & \multicolumn{6}{c}{Upper ScM} \\
\cline{1-6} \cline{7-12}
 \multicolumn{3}{c}{$E_L$ (MeV)} & \multicolumn{3}{c}{$B_L$ ($\mu_N^2$) } &
 \multicolumn{3}{c}{$E_U$ (MeV)} & \multicolumn{3}{c}{$B_U$ ($\mu_N^2$) } \\
 \cline{1-3} \cline{4-6} \cline{7-9} \cline{10-12}
 WFM &  Ref.~\cite{Shizuma26}   & Ref.~\cite{Maser96} &
 WFM &  Ref.~\cite{Shizuma26}   & Ref.~\cite{Maser96} &
 WFM &  Ref.~\cite{Shizuma26}   & Ref.~\cite{Maser96} &
 WFM &  Ref.~\cite{Shizuma26}   & Ref.~\cite{Maser96} \\
\cline{1-3} \cline{4-6} \cline{7-9} \cline{10-12}
2.83 & 2.59(12) & 2.59 &    2.05 & 1.23(4) & 1.11(27) &
3.54 & 3.16(11) & 3.25 &    1.57 & 1.45(4) & 1.86(46)\\
\end{tabular}\end{ruledtabular}
\label{tab2}
\end{table*}


\end{widetext}
{\it Conclusion.}

The structure of the scissors mode in $^{166}$Er is analyzed within the framework of WFM method. The results of calculations rather reasonably reproduce
the main features of the spectrum of 1$^+$ excitations in
the energy interval $2.4 - 3.5$ MeV (scissors mode region) observed in NRF experiments~\cite{Shizuma26,Maser96}. Such spectrum in $^{166}$Er shows a clear splitting into two bumps.
Similar splitting is also observed in a number of other nuclei of rare earths and in actinides (see papers~\cite{BaMoPRC18,BaMoPRC22,BaMoIJMP24}, where this issue was carefully scrutinized).

In the paper~\cite{Shizuma26} it was suggested that the observed splitting is explained by "nuclear triaxiality" ($\gamma$-deformation). However, the analysis of the observed splitting of the scissors mode in $^{232}$Th, carried out in paper~\cite{Oslo}, clearly showed that the assumption of non-axiality does not provides a satisfactory description of the $M1$ strength distribution.
Calculations performed within the WFM method allow to conclude that the discussed
splitting occurs due to the energy separation of conventional scissors and spin-scissors.

To understand the microscopic structure of the 1$^+$ states the authors of Ref.~\cite{Shizuma26} performed "large-scale shell-model calculations`` using the QVSM. As a result they have found two prominent peaks at 2.48 MeV and 3.18 MeV with $B(M1)\uparrow$ values of 2.0 $\mu_N^2$ and 0.86 $\mu_N^2$ respectively. The analysis of their spin and orbital components has shown that the lower peak has the scissorslike (orbital) character, whereas the higher peak is the spin-flip excitation. The authors were careful enough to make "a direct comparison" of these results with their experimental data.

The WFM theory predicts two states at 2.83 MeV and 3.54 MeV with $M1$ strengths of 2.05 $\mu_N^2$ and 1.57 $\mu_N^2$ respectively.
Analysis of the currents~\cite{BaMoPRC22} has shown that the state with an energy of 2.83 MeV is predominantly of the spin nature,  while the state at energy of 3.54 MeV is predominantly of the orbital one. The studying of the interplay of orbital and spin contributions to $B(M1)$ leads to the same conclusion.
So, our calculations do not support the interpretation proposed in paper~\cite{Shizuma26}.


\end{document}